\newcommand{\CLASSINPUTtoptextmargin}{19mm}%
\newcommand{\CLASSINPUTbottomtextmargin}{43mm}%
\newcommand{\CLASSINPUTinnersidemargin}{12.9mm}%
\newcommand{\CLASSINPUToutersidemargin}{12.9mm}%
\documentclass[conference,10pt,a4paper]{IEEEtran}
\usepackage{amsmath,amssymb,amsfonts,enumitem,,xfp}
\usepackage{soul}
\usepackage{color}
\usepackage{url}
\usepackage{cite}
\usepackage[detect-all]{siunitx}
\usepackage{tabularx,booktabs}
\usepackage{algorithmic}
\usepackage{graphicx}
\usepackage{pgfplots}
\usepackage{tikzscale}
\usepackage{tkz-euclide}
\usepackage{datatool}
\usepackage{textcomp}
\usepackage{xcolor}
\usepackage[siunitx]{circuitikz}

\ifCLASSOPTIONcompsoc
\usepackage[caption=false,font=normalsize,labelfont=sf,textfont=sf]{subfig}
\else
\usepackage[caption=false,font=footnotesize]{subfig}
\fi

\usetikzlibrary{external}
\usetikzlibrary{math} 
\usetikzlibrary{shapes.multipart}
\usetikzlibrary{shapes.geometric,calc,angles,positioning,intersections,quotes,decorations,babel,patterns,fit}
\usetikzlibrary{shapes.misc}

\tikzset{cross/.style={cross out, draw=black, minimum size=2*(#1-\pgflinewidth), inner sep=0pt, outer sep=0pt},
	cross/.default={1pt}}
\usepgfplotslibrary{external} 
\pgfplotsset{compat = newest}
\usepgfmodule{nonlineartransformations}

\DeclareMathOperator{\sinc}{sinc}

\def\BibTeX{{\rm B\kern-.05em{\sc i\kern-.025em b}\kern-.08em
		T\kern-.1667em\lower.7ex\hbox{E}\kern-.125emX}}

\newcommand{\s}[1]{\mathrm{#1}}
\newcommand{\ph}[3][]{\varphi_\s{#2}^\s{#1}(#3)}  
\renewcommand{\j}{\s{j}}

\definecolor{myblue}{RGB}{0,114,189}
\definecolor{myred}{RGB}{217,83,25}
\definecolor{mypurp}{RGB}{126,47,142}
\definecolor{myyellow}{RGB}{237, 177, 32}
\definecolor{mygreen}{RGB}{77,172,48}
\definecolor{mygreen2}{RGB}{1, 129, 114}
\definecolor{myred2}{RGB}{255,83,83}

\makeatletter

\def\@maketitle{\newpage
\bgroup\par\addvspace{0.5\baselineskip}\centering%
\ifCLASSOPTIONtechnote
   {\bfseries\large\@IEEEcompsoconly{\sffamily}\@title\par}\vskip 1.3em{\lineskip .5em\@IEEEcompsoconly{\sffamily}\@author
   \@IEEEspecialpapernotice\par{\@IEEEcompsoconly{\vskip 1.5em\relax
   \@IEEEtitleabstractindextextbox{\@IEEEtitleabstractindextext}\par
   \hfill\@IEEEcompsocdiamondline\hfill\hbox{}\par}}}\relax
\else
   \vskip0.2em{\EuMWtitlesize\ifCLASSOPTIONtransmag\bfseries\LARGE\fi\@IEEEcompsoconly{\sffamily}\@IEEEcompsocconfonly{\normalfont\normalsize\vskip 2\@IEEEnormalsizeunitybaselineskip
   \bfseries\Large}\@title\par}\vskip1.0em\par
   \ifCLASSOPTIONconference%
      {\@IEEEspecialpapernotice\mbox{}\vskip\@IEEEauthorblockconfadjspace%
       \mbox{}\hfill\begin{@IEEEauthorhalign}\@author\end{@IEEEauthorhalign}\hfill\mbox{}\par}\relax
   \else
      \ifCLASSOPTIONpeerreviewca
         {\@IEEEcompsoconly{\sffamily}\@IEEEspecialpapernotice\mbox{}\vskip\@IEEEauthorblockconfadjspace%
          \mbox{}\hfill\begin{@IEEEauthorhalign}\@author\end{@IEEEauthorhalign}\hfill\mbox{}\par
          {\@IEEEcompsoconly{\vskip 1.5em\relax
           \@IEEEtitleabstractindextextbox{\@IEEEtitleabstractindextext}\par\hfill
           \@IEEEcompsocdiamondline\hfill\hbox{}\par}}}\relax
      \else
         \ifCLASSOPTIONtransmag
           {\@IEEEspecialpapernotice\mbox{}\vskip\@IEEEauthorblockconfadjspace%
            \mbox{}\hfill\begin{@IEEEauthorhalign}\@author\end{@IEEEauthorhalign}\hfill\mbox{}\par
           {\vspace{0.5\baselineskip}\relax\@IEEEtitleabstractindextextbox{\@IEEEtitleabstractindextext}\vspace{-1\baselineskip}\par}}\relax
         \else
           {\lineskip.5em\@IEEEcompsoconly{\sffamily}\sublargesize\@author\@IEEEspecialpapernotice\par
           {\@IEEEcompsoconly{\vskip 1.5em\relax
            \@IEEEtitleabstractindextextbox{\@IEEEtitleabstractindextext}\par\hfill
            \@IEEEcompsocdiamondline\hfill\hbox{}\par}}}\relax
         \fi
      \fi
   \fi
\fi\par\addvspace{0.0\baselineskip}\egroup}

\def\EuMWtitlesize{\@setfontsize{\EuMWtitlesize}{24}{24pt}}
\def\EuMWauthorsize{\@setfontsize{\EuMWauthorsize}{11}{11pt}}
\def\EuMWaffilsize{\@setfontsize{\EuMWaffilsize}{10}{10pt}}
\def\EuMWcaptionsize{\@setfontsize{\EuMWcaptionsize}{9}{10pt}}
\def\EuMWbibsize{\@setfontsize{\EuMWbibsize}{8}{10pt}}

\def\@IEEEauthorblockNstyle{\EuMWauthorsize\@IEEEcompsocnotconfonly{\sffamily}\@IEEEcompsocconfonly{\large}}
\def\@IEEEauthorblockAstyle{\EuMWaffilsize\@IEEEcompsocnotconfonly{\sffamily}\@IEEEcompsocconfonly{\itshape}\@IEEEcompsocconfonly{\large}}
\def\@IEEEauthordefaulttextstyle{\EuMWauthorsize\@IEEEcompsocnotconfonly{\sffamily}\sublargesize}

\def\thebibliography#1{\section*{\refname}%
    \addcontentsline{toc}{section}{\refname}%
    \EuMWbibsize\@IEEEcompsocconfonly{\small}\vskip 0.3\baselineskip plus 0.1\baselineskip minus 0.1\baselineskip
    \list{\@biblabel{\@arabic\c@enumiv}}%
    {\settowidth\labelwidth{\@biblabel{#1}}%
    \leftmargin\labelwidth
    \advance\leftmargin\labelsep\relax
    \itemsep \IEEEbibitemsep\relax
    \usecounter{enumiv}%
    \let\p@enumiv\@empty
    \renewcommand\theenumiv{\@arabic\c@enumiv}}%
    \let\@IEEElatexbibitem\bibitem%
    \def\bibitem{\@IEEEbibitemprefix\@IEEElatexbibitem}%
\def\newblock{\hskip .11em plus .33em minus .07em}%
\ifCLASSOPTIONtechnote\sloppy\clubpenalty4000\widowpenalty4000\interlinepenalty100%
\else\sloppy\clubpenalty4000\widowpenalty4000\interlinepenalty500\fi%
    \sfcode`\.=1000\relax}

\long\def\@makecaption#1#2{%
\ifx\@captype\@IEEEtablestring%
\par\@IEEEtabletopskipstrut
\else
\@IEEEfigurecaptionsepspace
\fi
\setbox\@tempboxa\hbox{\normalfont\footnotesize {#1.}\nobreakspace\nobreakspace #2}%
\ifdim \wd\@tempboxa >\hsize%
\setbox\@tempboxa\hbox{\normalfont\footnotesize {#1.}\nobreakspace\nobreakspace}%
\parbox[t]{\hsize}{\normalfont\footnotesize\noindent\unhbox\@tempboxa#2}%
\else
\ifCLASSOPTIONconference \hbox to\hsize{\normalfont\footnotesize\hfil\box\@tempboxa\hfil}%
\else \hbox to\hsize{\normalfont\footnotesize\box\@tempboxa\hfil}%
\fi\fi
\ifx\@captype\@IEEEtablestring%
\@IEEEtablecaptionsepspace
\else
\fi}

\newlength\tablecaptiontotableskip
\newlength\figuretocaptionskip
\def\@IEEEfigurecaptionsepspace{\vskip\figuretocaptionskip\relax}%
\def\@IEEEtablecaptionsepspace{\vskip\tablecaptiontotableskip\relax}%

\def\abstract{\normalfont%
\@IEEEabskeysecsize\bfseries\textit{\abstractname}\,\bfseries\textit{---}\,%
\@IEEEgobbleleadPARNLSP}%

\def\IEEEkeywords{\normalfont%
\@IEEEabskeysecsize\bfseries\textit{\IEEEkeywordsname}\,\bfseries\textit{---}\,%
\@IEEEgobbleleadPARNLSP}%
\def\endIEEEkeywords{\relax\vspace{0.67ex}%
\par\if@twocolumn\else\endquotation\fi%
\normalsize\normalfont}%

\def\@IEEEauthorblockNtopspace{0ex}
\def\@IEEEauthorblockAtopspace{1mm}
\def\tablename{Table}
\def\thetable{\arabic{table}}
\def\IEEEkeywordsname{Keywords}
\def\subsubsection{\@startsection{subsubsection}{3}{\z@}{1.5ex plus 1.5ex minus 0.5ex}%
{0.7ex plus .5ex minus 0ex}{\normalfont\normalsize\itshape}}%
\newlength{\CPheadmatchindent}%
\def\@seccntformat#1{\hbox to\CPheadmatchindent{\csname the#1dis\endcsname}\hskip 0.1em \relax}
\IEEEilabelindentA \parindent
\IEEEilabelindent \IEEEilabelindentA
\IEEEelabelindent \parindent
\IEEEdlabelindent \parindent
\IEEElabelindent \parindent
\makeatother

\usepackage{fancyhdr}
\fancypagestyle{empty}{fancy}

\begin{document}
\raggedbottom
%

%
\title{Quasi-Monostatic Antenna Displacement\\in Radar Target Simulation}

\author{\IEEEauthorblockN{Axel Diewald, Benjamin Nuss and Thomas Zwick}
	\IEEEauthorblockA{Karlsruhe Institute of Technology, Karlsruhe, Germany\\
		Email: axel.diewald@kit.edu}
}

\maketitle

\begin{abstract}
Radar target simulators (RTS) have recently drawn much attention in research and commercial development, as they are capable of performing over-the-air validation tests under laboratory conditions by generating virtual radar echoes that are perceived as targets by a radar under test (RuT). The estimated angle of arrival (AoA) of such a virtual target is determined by the physical position of the particular RTS channel that creates it, which must therefore be considered when planning the setup. A single channel employs two antennas, one for the reception and the other for the re-transmission of the incoming radar signal. The antennas are positioned close together, but still spatially separated, thus an RTS channel can be considered quasi-monostatic, which causes non-negligible inaccuracies in the angle simulation. In this paper, the authors examine the analytical implications of this systemic deficiency on the angle estimation, which provides support for the design and setup of angle-simulating RTS systems. The mathematical derivations developed are verified by measurement.
\end{abstract}
%

\section{Introduction}

\IEEEPARstart{I}{n recent} years, the development of advanced driver assistance systems (ADAS) and therefore autonomous driving has continued to evolve. To ensure safe operation of ADAS functions, extensive testing procedures must be performed to validate not only the function itself but also the sensing devices on which it relies. In addition to other sensors such as camera, lidar, and ultrasound, radar sensors play an important role in the environment sensing task and therefore should be tested in an integrated manner. However, performing such validation tests on the road requires a lot of effort, as distances of hundreds of millions of kilometers must be covered to ensure that the radar sensors work properly \cite{MGLW2015,S2017,KW2016}. In addition, these tests are not repeatable, as individual traffic situations cannot be reproduced exactly.

Radar target simulators (RTS) have recently attracted much attention from the scientific and commercial communities because of their ability to thoroughly test radar sensors in-place and under laboratory conditions \cite{GSGVABMPP2018, WMNLD2020,RS2021,KT2021}. Their operating principle is to deceive a radar under test (RuT) by creating an artificial environment comprising of virtual radar targets. In order to fulfill this task, RTS systems receive, modify and re-transmit the radar signal emitted by the RuT. The simulation of the virtual target's angle of arrival (AoA) is determined by the physical position of the RTS's antennas relative to the RuT \cite{RS2021,KT2021,9337461}. Since the antennas for reception and re-transmission of a single RTS channel are close to each other, but still spatially separated, the system can be considered as quasi-monostatic.

In this paper, the authors investigate the influence of the RTS receiver and transmitter displacement on the radar signal processing and the resulting angle estimation. In the following, the fundamental operation of radar target simulation and the associated signal model, which accounts for the quasi-monostatic character of the antenna distribution, will be explained. The model is intended to support the setup design and antenna arrangement development process of highly accurate RTS systems. Finally, a measurement campaign is presented, that verifies the signal model and serves as a reference for the simulation.


\section{Radar Target Simulation}

\begin{figure}[]
	\centering
	\includegraphics[width=1\columnwidth]{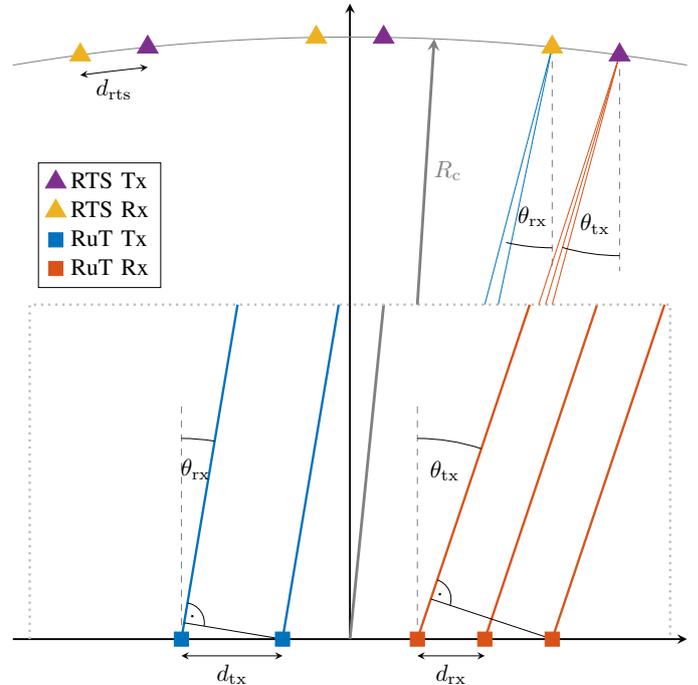}
	\caption{Conceptual representation of the quasi-monostatic characteristic of radar target simulation that leads to differences in signal path lengths and propagation delays. The scaling and spacing of the figure were chosen for clarity and are not representative.}
	\label{fig_qm_concept}
\end{figure}

Since detailed explanations of the general principle of operation of RTS systems can be found extensively in the literature \cite{GMS18,DMW18,s20092714}, only their basic operation will be described here. An RuT is placed in the center of a semi-circle formation with radius $R_\s{c} \approx \SI{1}{\meter}$, along which the RTS antenna pairs are distributed. Each pair consists of a receiver and a transmitter, which receive the signal emitted by the radar sensor and re-transmit it after the virtual target modifications have been applied. The principle is shown in Fig. \ref{fig_qm_concept}. For the generation of the virtual targets the signal is down converted to a lower intermediate frequency $f_\s{rts}$ before a delay, frequency shift and attenuation are applied, which translate to the target's range, Doppler shift and radar cross section (RCS). Subsequently, it is up converted back to its original carrier frequency $f_\mathrm{c}$ and re-transmitted towards the RuT. Variable AoAs are realized either through individual, mechanically movable RTS antenna pairs \cite{GR2017,KC20xx,9337461} or by an equidistant distribution of pairs and electronic switching between them \cite{GMS18,RS2021,KT2021}.

Due to the aforementioned spatial separation between the RTS receive and transmit antenna, the radar signal is received and re-transmitted at slightly different angular positions from the RuT's point of view. This causes the detected angle to deviate from the physical position of the RTS receiver as well as from that of the transmitter. Considering a RuT that resolves the AoA only in one dimension, this inaccuracy can be circumvented by arranging receiver and transmitter perpendicular to the resolvable domain. However, since modern and future radar sensors are capable of resolving targets in both the azimuth and elevation plane simultaneously, this technique does not resolve the issue and the uncertainty in the angle simulation remains.


\section{Signal Model}

In order analyze the effects of the RTS's quasi-monostatic antenna displacement on the radar's angle estimation, the radar signal that is transmitted by the RuT and modified by the RTS will be modeled. Since conventional radars primarily use the signal phase to estimate the angle of arrival of detected targets, the following equations are derived with respect to it.

For simplicity, a frequency-modulated continuous wave (FMCW) radar is assumed, whose $N_\s{tx}$ transmit and $N_\s{rx}$ receive antennas are distributed only in the azimuth plane. Its signal phase can be described in terms of time $t \in [0,T]$ as follows
\begin{align}
	\ph{tx}{t} &= 2\pi \int_0^{t} \left(f_\s{c} + \frac{B}{T} t'\right) dt' = 2\pi \left[f_\s{c} t + \frac{B}{2T} t^2\right]
\end{align} 
where $T$ describes the chirp period, $f_\s{c}$ the radar signal's start frequency and $B$  the signal's bandwidth. After propagating through free space and being received, modified and re-transmitted by the RTS, the signal is received by the radar and mixed with the original transmit signal to form the beat signal. It is then discretized by the RuT's analog-to-digital converter (ADC) and a discrete Fourier transform (DFT) and subsequent detection is performed. After that the signal and its phase can be expressed as follows
\begin{align}
	x_\s{R}[f_\s{R}] =& \ N_\s{s} \cdot \exp \left\{ \j \varphi_\s{R}[f_\s{R}] \right\} \label{eq_x_range} \\
	\varphi_\s{R}[f_\s{R}] =& \ 2\pi \left[f_\s{c} \tau_\s{c} + f_\s{rts} \tau_\s{rts} + \frac{1}{2} \left( B \tau -  f_\s{R} \right)\right]
\end{align}
where $N_\s{s}$ is the number of samples, $f_\s{R}$ the DFT bin index, $\tau_\s{c}$ the free space propagation delay, $\tau_\s{rts}$ the RTS's time delay and $\tau = \tau_\s{c} + \tau_\s{rts}$ the total time delay between the receive and transmit signal. A more detailed derivation of the intermediate steps can be found in \cite{9525811}. The free space propagation delay $ \tau_\s{c} = \tau_\s{tx} + \tau_\s{rx}$ combines the propagation delays of the transmitted signal from the RuT to the RTS $\tau_\s{tx}$ and that of the returning signal from the RTS to the RuT $\tau_\s{rx}$, which in turn can be expressed as
\begin{align}
	\tau_\s{tx} &= \frac{R_\s{c} + d_\s{tx} \cdot n_\s{tx} \cdot \sin(\theta_\s{rx})}{c_\s{0}} \\
	\tau_\s{rx} &= \frac{R_\s{c} + d_\s{rx} \cdot n_\s{rx} \cdot \sin(\theta_\s{tx})}{c_\s{0}}
\end{align}
where $R_\s{c}$ denotes the physical distance between the RuT and the RTS front ends. $\theta_\s{rx/tx} \in [\SI{-90}{\degree},\SI{90}{\degree}]$ describes the azimuth angle of the RTS receive/transmit front end, as seen by the RuT, $d_\s{tx/rx}$ the element spacing of the RuT's transmit/receive antennas and $n_\s{tx/rx} \in\allowbreak[0,\ldots,N_\s{tx/rx}-1]$ indexes the antenna elements. The above equations hold only if the far-field condition is satisfied by the setup \cite{longhurst1986geometrical}. After the range detection ($f_\s{R} = B\tau$), beamforming is applied to estimate the target's AoA
\begin{align}
	x_{\s{A}}[\alpha] =& \sum_{n_\s{tx}=0}^{N_\s{tx}-1} \sum_{n_\s{rx}=0}^{N_\s{rx}-1} x_{\s{R}} \left[ n_\s{tx},n_\s{rx} \right] \nonumber \\
	& \exp \left\{-\j2\pi\frac{\left(d_\s{tx} \cdot n_\s{tx} + d_\s{rx} \cdot n_\s{rx}\right) \cdot \sin \left(\alpha\right)}{\lambda} \right\}
\end{align}
where $\lambda$ is the radar signal's free space wavelength and $\alpha \in [\SI{-90}{\degree},\SI{90}{\degree}]$ is orientated equal to $\theta_\s{rx/tx}$. Utilizing the partial sum of a geometric series \cite{Bron01} and ${\sin(x) \approx x}$ for ${|x| \ll 1}$ the expression can be simplified to
\begin{align}
	x_{\s{A}}[\alpha] =& \; A N_\s{s} N_\s{tx} N_\s{rx} \cdot \exp \left\{ \j \varphi_\s{A} \right\} \nonumber \\
	& \sinc \left( \frac{d_\s{tx} \cdot N_\s{tx}}{\lambda} \cdot \left( \sin(\theta_\s{rx}) - \sin \left(\alpha\right) \right) \right) \nonumber \\
	& \sinc \left( \frac{d_\s{rx} \cdot N_\s{rx}}{\lambda} \cdot \left( \sin(\theta_\s{tx}) - \sin \left(\alpha\right) \right) \right) \label{x_a_final}\\
	\varphi_\s{A} =& \; 2\pi \left[  \left( f_\s{c} + \frac{B}{2} \right) \frac{2 R_\s{c}}{c_\s{0}} + \left( f_\s{rts} + \frac{B}{2} \right) \tau_\s{rts} \right. \nonumber \\
	& \qquad +  \frac{d_\s{tx}}{2\lambda} \cdot (N_\s{tx}-1) \cdot \sin(\theta_\s{rx}) \nonumber \\
	& \qquad \left. + \frac{d_\s{rx}}{2\lambda} \cdot (N_\s{rx}-1) \cdot \sin(\theta_\s{tx}) \right] \label{ph_a_final}
\end{align}
The maximum of $x_{\s{A}}[\alpha]$ yields the detected angle, that resides between the physical positions of RTS receiver and transmitter, whose spacing can be approximated by
\begin{align}
	d_\s{rts} \approx R_\s{c} \cdot \left[\sin(\theta_\s{tx}) - \sin(\theta_\s{rx}) \right]
\end{align}
As can be deducted from \eqref{x_a_final} and \eqref{ph_a_final}, the RuT's antenna geometry has an impact on the detected angle and must therefore be taken into account for precise and angle-accurate radar target simulation.


\section{Measurement}

\begin{figure}[]
	\centering
	\tikzmath{
\trxw = .75;
\trxh = \trxw/2 * sqrt(3);
\lmw = 6;
\lmh = 1;
\lmx = 3;
\lmy = 4;
\rmx = \lmx + \lmw/3;
\rmy = \lmy + .0;
\rmw = 1;
\rmh = \lmh;
\rmmx = \rmx + \rmw/2;
\rmmy = \rmy + \rmh/2;
\rmr = \rmh / 2 - .05;
\rmra = \rmr+.2;
\rmas = -45;
\rmae = 60;
\txx = \rmx + \rmw/2 - \trxw/2;
\txy = \rmmy - \trxh/2;
\rxx = 2;
\rxy = \txy;
\rw = 2;
\rh = 1;
\rx = \lmx - \rw/2;
\ry = 0;
\rlx = \rx+\rw/2;
\rly = \ry+\rh/2;
}

\begin{tikzpicture}
	
	\draw[fill=mygreen,very thick] (\rx,\ry) -- ++(\rw,0) -- ++(0,\rh) -- ++(-\rw,0) -- cycle;
	\draw[white] (\rlx,\rly) node[align=center,anchor=center] {RuT};

	\draw[fill=lightgray,thick] (\lmx,\lmy) -- ++(\lmw,0) -- ++(0,\lmh) -- ++(-\lmw,0) -- cycle;
	\draw[fill=white,thick] (\lmx+.1,\lmy+.1) -- ++(\lmw-.2,0) -- ++(0,\lmh-.2) -- ++(-\lmw+.2,0) -- cycle;
	\draw[fill=gray,thick]  (\lmx+.1,\lmy+.4) -- ++(\lmw-.2,0) -- ++(0,.2) -- ++(-\lmw+.2,0) -- cycle; 
	\draw[gray,thick,latex-latex] (\rmx+\rmw+.5,\lmy+.25) -- ++(2,0);
	\draw[gray] (\lmx+\lmw-.25,\lmy-.2) node[align=center,anchor=east] {Linear motor};
	
	\draw[fill=gray, thick] (\rmx,\rmy) -- ++(\rmw,0) -- ++(0,\rmh) -- ++(-\rmw,0) -- cycle;
	\draw[fill=teal!50, thick] (\rmmx,\rmmy) circle (\rmr);
	\draw[] (\rmmx,\rmmy) node[cross=9.25pt,rotate=45-25,white]{};
	\draw[teal!50,thick,latex-latex] (\rmmx,\rmmy)++(\rmas:\rmra) arc [start angle=\rmas, end angle=\rmae,radius=\rmra];
	\draw[teal!50] (\rmx+.5,\lmy+\lmh+.25) node[align=center,anchor=west] {Rotation motor};

	\draw[fill=mypurp,very thick,rotate around={-25:(\txx+\trxw/2,\txy+\trxh/2)}] (\txx,\txy) -- ++(\trxw,0) -- ++({-\trxw/2},\trxh) -- cycle;
	\draw[mypurp] ((\txx+\trxw/2+.1,\txy-.1) node[align=center,anchor=north] {Tx};

	\draw[fill=myyellow,very thick,rotate around={5:(\rxx+\trxw/2,\rxy+\trxh/2)}] (\rxx,\rxy) -- ++(\trxw,0) -- ++({-\trxw/2},\trxh) -- cycle;
	\draw[myyellow] ((\rxx+\trxw/2-.3,\txy-.05) node[align=center,anchor=north] {Rx};

	\draw[black,<->,thick] (\rxx+\trxw/2,\rxy+1) -- (\txx+\trxw/2,\rxy+1);
	\draw[black] ({(\rxx+\txx+\trxw)/2},\rxy+1) node[align=center,anchor=south] {$d_\mathrm{rts}$};
	
	\draw[black,<->,thick] (\txx+1,\ry+\rh) -- (\txx+1,\lmy);
	\draw[gray,dashed] (\rx+\rw,\ry+\rh) -- (\txx+1.5,\ry+\rh);
	\draw[black] (\txx+1,\lmy-2) node[align=left,anchor=west] {$R_\mathrm{c}$};

	\draw[myblue,-Latex,very thick] (\rx+\rw/2-.1,\ry+\rh+.25) --  (\rxx+\trxw/2+.1,\rxy-.25);
	\draw[myred,Latex-,very thick] (\rx+\rw/2+.1,\ry+\rh+.25) --  (\txx+\trxw/2-.25,\txy-.25);




\end{tikzpicture}
	\caption{Concept of the measurement setup in top view.}
	\label{fig_meas_setup}
\end{figure}
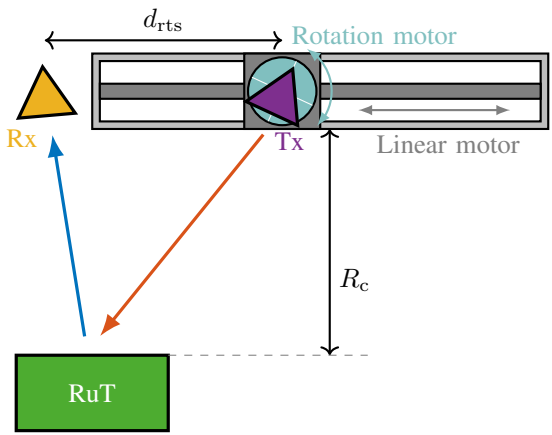

\begin{figure}[]
	\centering
	\input{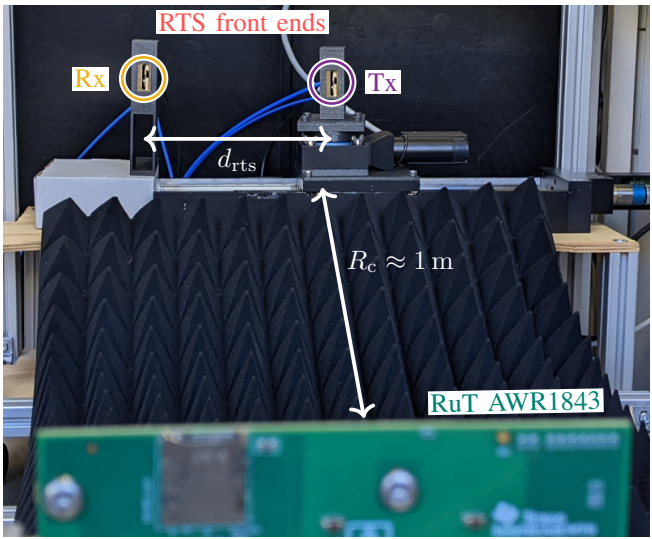}
	\caption{Photograph of the measurement setup.}
	\label{fig_foto_meas_setup}
\end{figure}


For the verification of the mathematical derivations previously developed, a measurement campaign was conducted. For this, a digital RTS system, whose individual components are presented in \cite{vehicles3020016}, was used. The back end of the RTS was realized with an UltraScale+ RFSoC FPGA from Xilinx. The RTS's intermediate frequency was set to $f_\s{rts} = \SI{500}{\mega\hertz}$. As the RuT a Texas Instruments AWR1843BOOST radar board was employed. It was configured to use $N_\s{tx} = 2$ transmit and $N_\s{rx} = 4$ receive antennas, with an element spacing of $d_\s{tx} = 2\lambda$ and $d_\s{rx} = \frac{\lambda}{2}$, respectively. The bandwidth was set to $B = \SI{1}{\giga\hertz}$ and the start frequency to $f_\s{c} = \SI{77}{\giga\hertz}$.

The concept of the measurement setup can be seen in Fig. \ref{fig_meas_setup}. The RuT was placed at distance $R_\s{c}$ in front of a single RTS receiver and transmitter. Multiple measurements were performed in between which the spatial separation of the RTS antennas was incrementally increased. For this, a linear motor moved the RTS transmitter laterally while the receiver's position was fixed. An additional rotation motor, that was mounted on the linear motor, compensated for the change in the relative angle of the RuT as seen by the RTS transmitter. This was necessary in order to neutralize any effect that the RTS transmitter's directional antenna pattern might have on the radar processing chain. The introduced range offset was accounted for in the simulation, but remained smaller than the range resolution of the radar. Fig. \ref{fig_foto_meas_setup} shows a photograph of the measurement setup.

\begin{figure}[]
	\centering
	\includegraphics[width=1\columnwidth]{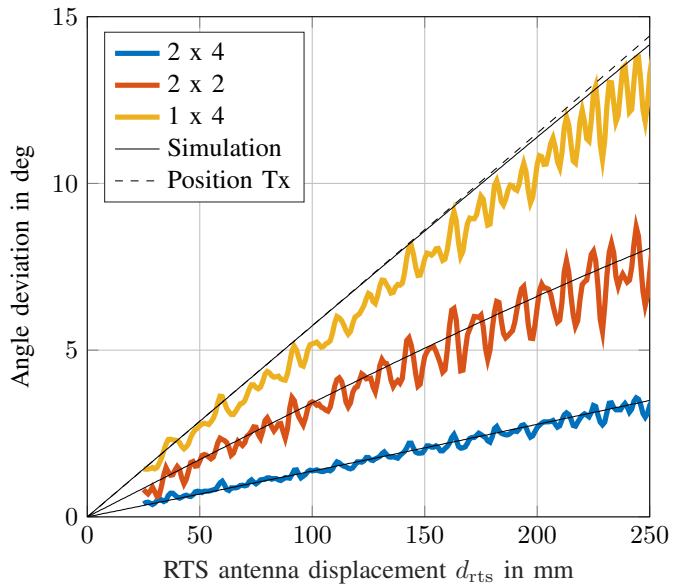}
	\caption{Measurement and simulation results for the angle deviation in relation to the RTS antenna displacement.}
	\label{fig_meas_sim_ang_dev}
\end{figure}

The returning radar signal received by the RuT was processed to estimate the AoA. This was done by applying a DFT and subsequent range detection to the beat signal. Thereupon, beamforming was employed to determine the AoA. Fig. \ref{fig_meas_sim_ang_dev} plots the deviation of the detected angle from the angular position of the RTS receiver over the RTS displacement $d_\s{rts}$. The same measurement data was processed with different antenna selections (2x4, 2x2, 1x4) to showcase the influence of the antenna geometry on the detected angle. As can be observed, simulation and measurement show good agreement, which verifies the analytical expressions previously derived. The fluctuations in the measurement curves are due to multipath reflections caused by the mechanical structure of the setup. The position of the RTS transmitter mostly matches the detected angle of the 1x4 antenna configuration, which makes sense since this configuration only has differences in the signal path lengths for the returning signal from the RTS transmitter to the RuT.

\section{Conclusion}

In this paper, the authors investigate the influence of the quasi-monostatic characteristic of radar target simulation (RTS) transceivers on the angle of arrival (AoA) estimation of a potential radar under test (RuT) in regard to deviations between detected AoA and physical position of the transceiver. For this purpose, an analytical signal model was derived and verified by measurement. The knowledge obtained can support the antenna arrangement process when designing angle-simulating RTS systems.


\section*{Acknowledgment}

The authors would like to thank PKTEC GmbH, Schutterwald, Germany, for providing the front-end transceiver hardware and Texas Instruments Inc., Dallas, TX, USA, for supplying the radar under test (RuT).


\bibliographystyle{IEEEtran}

\bibliography{literature}

\end{document}